\documentstyle[12pt]{article}
\makeatletter
\ifcase \@ptsize
\or 
\or 
\fi \advance\textheight by \topskip

\ifcase \@ptsize
\or 
\or 
\fi

\begin{document}

\centerline{\Large\bf Boundary scattering in the principal chiral
model\footnote{ Based on a talk presented by NM at the S\~ao Paulo
workshop on `Integrable field theories, solitons and duality, July
2002}}

\vskip 0.2in
 \centerline{\large Niall MacKay\footnote{\tt
nm15@york.ac.uk} and Ben Short\footnote{bjs108@york.ac.uk}}

\vskip 0.1in \centerline{\em Department of Mathematics, University
of York, York YO10 5DD, UK }

 \vskip 0.1in \abstract{In recent work on the ($G\!\times
\!G$-invariant) principal chiral model with boundary, we found
that both classically integrable boundary conditions and quantum
boundary $S$-matrices were classified by the symmetric spaces
$G/H$. The connection is explained by the presence of a `twisted
Yangian' algebra of non-local charges.}

\newcommand{\be}{\begin{equation}}
\newcommand{\ee}{\end{equation}}
\newcommand{\bea}{\begin{eqnarray}}
\newcommand{\eea}{\end{eqnarray}}
\newcommand{\beaa}{\begin{eqnarray*}}
\newcommand{\eeaa}{\end{eqnarray*}}
\newcommand{\g}{{\bf g}}
\newcommand{\h}{{\bf h}}
\baselineskip 18pt
\parskip 10pt
\parindent 10pt

\section{Introduction}

In a recent paper \cite{[1]}, we investigated the boundary
principal chiral model, defined by\be\label{pcmlagr} {\cal L} =
{1\over 2}{\rm Tr}\left(
\partial_\mu g^{-1}
 \partial^\mu g\right)
\ee where the field $g(x,t)$ takes values in a compact classical
Lie group $G$, defined on Minkowski space with $-\infty<x\leq0$.

We found connections with symmetric spaces $G/H$ in two ways.
First, we discovered mixed boundary conditions which allowed
conservation and commutation of the local charges essential to
integrability. In these, the field $g$ was restricted at $x=0$ to
lie either in $H$ or in the Cartan immersion of $G/H$ (an
embedding of $G/H$ as a submanifold of $G$) or in a translate of
one of these. We describe this briefly in section two. Next, we
investigated rational solutions of the boundary Yang-Baxter or
`reflection' equation (which would be needed for the construction
of boundary $S$-matrices), and found that those of the simplest
form were also in correspondence with the symmetric spaces, there
being solutions parametrized by the Cartan immersion of $G/H$. The
beauty of this second correspondence in particular is easily
swamped by the exhaustive (and exhausting) detail needed to prove
it case-by-case, and in section three we instead give a flavour of
how it works. The connection between the two is clarified by the
discovery that a remnant of the bulk Yangian algebra of non-local
charges survives on the half-line. This consists of the generators
of the $H$ symmetry expected, along with a set of non-linear
combinations of non-local charges, and together these form a
twisted Yangian. We describe this in section four.

\section{Classical boundary conditions}

First, the classical boundary conditions (BCs). The boundary
equation of motion is$$ {\rm Tr}(g^{-1} \partial_1 g.g^{-1}\delta
g)=0 \qquad {\rm at} \;\;\; x=0\,,$$ where the variation is over
all $\delta g$ such that $g^{-1}\delta g\in\g$, the Lie algebra of
$G$ . This is clearly satisfied by the Neumann BC $\partial_1
g=0$, or by the Dirichlet condition $\delta g=0$. But it is also
satisfied by two types of mixed condition,\bea\label{bc2}& {\rm
chiral}\hspace{0.3in} & g(0)\in g_L H
g_R^{-1}\hspace{0.3in}\\[0.1in] \label{bc3} {\rm and}&{\rm
non-chiral}\hspace{0.3in}& g(0)\in g_L {G\over H}\,
g_R^{-1}\,,\hspace{0.3in} \eea where $G/H
\cong\{\alpha(g)g^{-1}|g\in G\}$ is the Cartan immersion of $G/H$
in $G$, with $H$ the subgroup of $G$ invariant under an involutive
({\em i.e.\ }$\alpha^2=1$) automorphism $\alpha$, while $g_L,g_R$
are fixed elements of $G$. (Actually, one has to be a little
careful; the Cartan immersion is a local diffeomorphism, but may
be (finitely) neither 1-1 nor onto, as explained in \cite{[5]}.)

This requirement that $g$ be restricted to lie in such a
`D-submanifold' is then supplemented by a Neumann condition on it.
To see this, define the $\g$-valued conserved currents which
generate the $G\!\times\!G$ symmetry $g\mapsto UgV^{-1}$,
\be\label{lrcurr} j(x,t)_\mu^L=\partial_{\mu} g \,g^{-1} , \qquad
j(x,t)_\mu^R = - g^{-1}\partial_{\mu} g\,, \ee and notice that the
boundary equation of motion is equivalent to ${\rm Tr}(j_0 j_1)=0$
at $x=0$ (using either $j^L$ or $j^R$ -- it doesn't matter which
because of cyclicity of trace). Our chiral  BC (\ref{bc2}) is
equivalent to\be g_L^{-1} j_0^L g_L = \alpha ( g_L^{-1} j_0^L g_L
)\in\h \hspace{0.3in} {\rm and} \hspace{0.3in}g_R^{-1} j_1^L g_R =
\alpha ( g_R^{-1} j_1^L g_R )\in\h\ee at $x=0$, where
$\g=\h\oplus{\bf k}$; $\h$ generates $H$ and is the $+1$
eigenspace, while ${\bf k}$ is the $-1$ eigenspace. The boundary
equation of motion then requires that the space components take
values in ${\bf k}$. Together these give \be g_L^{-1} j_+^L g_L =
\alpha ( g_L^{-1} j_-^L g_L )\hspace{0.3in}{\rm at}\;\;x=0\,,\ee
and similarly for $j^R$. It is an easy exercise to show similarly
that the non-chiral BC (\ref{bc3}) implies $g_L^{-1} j_0^L
g_L=\alpha(g_R^{-1}j_0^R g_R)$ and so requires\be g_L^{-1} j_+^L
g_L = \alpha ( g_R^{-1} j_-^R g_R )\hspace{0.3in}{\rm
at}\;\;x=0\,.\ee

The conserved, commuting local charges of the model are built up
using certain special choices of invariant tensors \cite{[2]}.
That they remain conserved and commuting on the half-line in the
presence of these BCs follows from the simple behaviour of these
tensors under $\alpha$. When $\alpha$ is an inner automorphism,
the tensors are of course invariant under $\alpha$, but when
$\alpha$ is outer they may not be so. In fact they always have
eigenvalue $\pm 1$ under $\alpha$ \cite{[3]}, and this is enough
to ensure that one conserved charge may be constructed from each
$\pm$ spin pair of local charges -- for example, energy but not
momentum. This, and the consistency of the Poisson brackets with
these BCs, is described in detail in \cite{[1]}.

The final point concerns the moduli space of parameters
$(g_L,g_R)$ by which we may `twist' our BCs. For the chiral BCs,
it is clear that this will be $G/H\times G/H$, modulo some finite
set. For each such BC we must construct a boundary $S$-matrix, and
this should respect the remnant $g_L H g_L^{-1} \times g_R H
g_R^{-1}$ of the $G\!\times\!G$ symmetry.

\section{Boundary $S$-matrices}

Assuming that these results survive quantization, we must now seek
solutions of the reflection equation from which to construct
boundary $S$-matrices. Recall the reflection equation
 \be S(\theta-\phi).\, I \otimes
K(\theta).\, S(\theta+\phi).\,I \otimes K(\phi)=
 I \otimes K(\phi).\,S(\theta+\phi)
.\,I \otimes K(\theta).\, S(\theta-\phi)\,, \ee acting on
$V\otimes V$ where $V$ is the defining (vector) representation (or
possibly, for $SU(N)$, its conjugate) of a classical  group $G$.
(The corresponding calculations for any exceptional groups are a
rather tougher proposition.) We begin with the known rational
$S:V\otimes V\rightarrow V\otimes V$, and must find
$K:V\rightarrow V$. To do so we make the ansatz \be\label{form}
K(\theta)=K_1(\theta)\equiv\rho(\theta)E \qquad{\rm or}\qquad
K(\theta)=K_2(\theta)\equiv\frac{\tau(\theta)}{1-
c\theta}(I+c\theta E)\,, \ee where $c$ and $E$ are constants to be
determined, the latter an $N\!\times\! N$ matrix, and
$\rho(\theta)$ and $\tau(\theta)$ are scalar prefactors,
undetermined by the reflection equation but which are needed to
satisfy the unitarity and crossing-unitarity conditions (which we
do not detail here) required of boundary $S$-matrices. The second
form $K_2$ is basically Cherednik's ansatz \cite{[4]}. Physically,
these are $S$-matrices with at most two `channels'; $K_1(\theta)$
will have no non-trivial pole structure, while with $K_2(\theta)$
there is the possibility of precisely one non-trivial boundary
bound state.

On substituting these into the reflection, unitarity and
crossing-unitarity conditions, we obtain various sets of
conditions on $E$ (and $c$). Our result is that each set
corresponds to a symmetric space $G/H$, and that in each case the
space of $E$ which satisfies the conditions is isomorphic to
(sometimes a finite multiple of) a translate of the Cartan
immersion $G/H\hookrightarrow G$.

For a brief indication of how such results are arrived at, let us
look at the example of $SU(N)$. There are two forms of the
reflection equation to consider: that acting on $V\otimes V$, and
that on $V\otimes \bar{V}$. For $V\otimes V$, one finds that $K_2$
solves all the conditions if \be\label{c1} E^\dag E=I,\qquad
E=E^\dag \qquad {\rm and} \qquad c=-{2N\over i\pi{\rm Tr}E}.\ee
The only $K_1$-type solution is really the degeneration of this
when Tr$E=0$. For $V\otimes \bar{V}$, one finds that $K_2$ gives
no solutions, but $K_1$ works with \be \label{c2} E^\dag
E=I,\qquad {\rm det} E=\pm 1 \qquad {\rm and}\qquad E=\pm E^T\ee
(the two choices of sign being independent).

Now consider $\alpha: SU(N)\rightarrow SU(N)$, $U\mapsto XUX$
where $X= $diag$(+1$ $M$ times, $-1$ $N-M$ times$)$. Then
\be\label{Grass} {SU(N)\over S(U(M)\times U(N-M)} \cong \{
XUXU^{-1}X \vert U\in SU(N)\}\,,\ee a translation by $X$ (and thus
in $U(N)$ when $N-M$ is odd) of the Cartan immersion. It is easy
to see that this is contained in the space of $E$ satisfying
(\ref{c1}) (and with Tr$E=$Tr$X=2M-N=-4/c$), although to prove
that they are equivalent takes rather longer. (The $K_1$ solution
corresponds to the case $2M=N$.)

This covers the $V\otimes V$ solutions. Now for $V\otimes \bar{V}$
consider \be {SU(N)\over SO(N)}\cong \{ \bar{U}U^{-1}\vert U\in
SU(N)\}=\{E\in SU(N) \vert E=E^T\}\ee -- so that here the set of
$E$ satisfying (\ref{c2}) is a two-fold copy of the symmetric
space. Note again that it is simple to check that the first set is
contained in the second, rather harder to check the reverse.
Finally, for $N$ even, the choice $E=-E^T$ corresponds to
$SU(N)/Sp(N)$.

We can now construct the boundary $S$-matrices for the principal
chiral model in the form \be Y(\theta)\Big(K_L(\theta)\otimes
K_R(\theta)\Big) \ee where $K_{L,R}(\theta)$ are $L$ and $R$
copies of the same type of $K$ as found above, with their scalar
prefactors constructed so as to give no poles on the physical
strip $0\leq$ Im $\theta\leq i\pi/2$. The CDD factor $Y$ is then
used to implement the pole structure we desire.

Let us note some facts about the $K_2$-type solution for the
Grassmannian symmetric space (\ref{Grass}). First, we might expect
that it, like (\ref{Grass}), would be invariant under $M\mapsto
N-M$, and indeed it is, since (after a relabelling of bases)
$X\mapsto -X$, $E\mapsto -E$ and $c\mapsto -c$, leaving $I+c\theta
E$ invariant; the scalar prefactors can then be constructed to
respect this. Second, the boundary $S$-matrix is, like the
classical BCs, parametrized by $G/H\times G/H$ (again modulo a
finite set), and commutes with  $g_L H g_L^{-1} \times g_R H
g_R^{-1}$ if we choose $E_L=g_L X g_L^{-1}$ (and similarly $E_R$).

We choose this $K_2$ to have a pole at $1/c$, since \be\label{S}
{1\over1- c\theta}(I+c\theta E) = P_- + { {N-2M\over 2N} +
{\theta\over i\pi} \over {N-2M\over 2N} - {\theta\over i\pi} }
P_+\;,\hspace{0.3in} P_\pm = {1\over 2}(I\pm E)\,,\ee projects at
this value onto the restriction to the $SU(M)$ subspace. If the
bulk particle has mass $m_1$, this gives a boundary bound state
(BBS) of mass $m_1\sin\left({2M\over N}\right)$. This is the
starting point of a bootstrap programme in which we can go on to
construct the scattering of all the bulk particles (of which there
are $N-1$, with $m_a=m \sin{a\pi\over N}$) off the boundary ground
state, and to seek higher BBSs. Finally, of course, we can also
scatter the bulk particles off the BBSs themselves. In this
context our initial ansatz of a maximum of one pole seems rather
natural -- it is the scattering of the higher bulk and boundary
states which will have more poles.

\section{Twisted Yangian non-local charges}

The connection between sections two and three is provided by
examining the non-local charges which survive on the half-line
with these boundary conditions.

In the bulk model, the $G\times G$ symmetry sits inside a larger
$Y(\g)\times Y(\g)$ symmetry, where $Y(\g)$ is the Yangian
algebra. This is generated by charges\bea \label{Q0} Q^{(0)a} & =
& \int j_0^a \,dx\\ \label{Q1} Q^{(1)a} & =  & \int j_{1}^{a} {dx}
- {1\over 2}f^a_{\;\;bc}\int j_{0}^{b}(x) \int^{x} j_{0}^{c}(y)
\,{dy} \,{dx}\,\eea using $j^L$ and $j^R$ respectively, decomposed
into $j=j^a t_a$ where the $t^a$ are generators of $\g$ with
$[t_a,t_b]=f_{ab}^{\;\;\;c} t_c$. The integrals are over all
space, $(-\infty,\infty)$ for the bulk model. But on the half-line
$(-\infty,0]$, these charges are no longer generally conserved.
However, there are two important sets of charges which do remain
(classically) conserved. Writing $\h$-indices as $i,j,k,..$ and
${\bf k}$-indices as $p,q,r,...$, and noting that the only
non-zero structure constants are $f^i_{\;\;jk}$ and $f^i_{\;\;pq}$
(and cycles thereof), these are \bea\label{Q0b} &&Q^{(0)i}\\{\rm
and} && \widetilde{Q}^{(1)p} \equiv Q^{(1)p} + {1\over 2}
f^p_{\;\;qi} Q^{(0)i} Q^{(0)q}\,, \label{Q1b} \eea as described in
\cite{[8]}. The first set generates $H$, while the second is a
remnant of the non-local charges. After quantization, this second
set of charges may be written as \be\widetilde{Q}^{(1)p} =
Q^{(1)p} + {1\over 4} [C_2^\h, Q^{(0)p}] \label{Q1q}\,. \ee where
($\hbar=1$ and) $C_2^\h\equiv \gamma_{ij}Q^{(0)i} Q^{(0)j}$ is the
quadratic Casimir operator of $\g$ restricted to $\h$, with
$\gamma_{ij}=f_{ia}^{\;\;\;b}f_{jb}^{\;\;\;a}$. Conservation of
these charges can then be used to determine the boundary
$S$-matrices, up to an overall scalar factor. The procedure is
that, first, conservation of the $Q^{(0)i}$ requires that $K$ act
trivially on irreps of $H$ (hence the decomposition of (\ref{S})
into projectors $P_\pm$). Second, conservation of the
$\widetilde{Q}^{(1)p}$ determines the relative values of the
coefficients of these projectors in terms of the values of
$C_2^\h$. Details may be found in \cite{[8]} and for all $\g$ in
\cite{[9]}.

In the same way that $Y(\g)$ may be thought of as a deformation of
a polynomial algebra $\g[z]$ (with $Q_1^a=zQ_0^a$), the algebra
generated by the $Q^{(0)i}$ and $\widetilde{Q}^{(1)p}$ may be
viewed as a deformation of the subalgebra of (`twisted')
polynomials in $\g[z]$ invariant under the combined action of
$\sigma$ and $z\mapsto -z$. For this reason it is known as the
`twisted Yangian', and we write it $Y(\g,\h)$. It is well-known
and -studied in the literature (see for example \cite{[6]}),
although not in this general form.

\section{Final remarks}

The structures we have described form the foundation for a full
boundary bootstrap programme, to discover the boundary spectrum
and its interactions with the bulk. Whilst this will be very tough
to carry through, some progress has been made \cite{[10]}. There
are many other directions in which progress might be possible --
bulk $G/H$ sigma models, models with Wess-Zumino terms and models
based on supergroups come to mind -- but one of the most
intriguing is the apparent unification of various exact $S$-matrix
phenomena for exceptional groups by the Deligne series and the
magic square \cite{[9],[11]}.

\vskip 0.2in \noindent{\bf Acknowledgments}\\ NM would like to
thank the IFT, Sao Paulo for its hospitality, and the Royal
Society for a conference/visit grant. BS thanks the UK EPSRC for a
Ph.D.\ studentship.

\end{document}